# A comparative computational study of the electronic properties of planar and buckled silicene

**Harihar Behera[1] and Gautam Mukhopadhyay[2]**

*Indian Institute of Technology Bombay, Powai, Mumbai-400076, India*

[1]*Email: harihar@phy.iitb.ac.in; behera.hh@gmail.com*
[2]*Email: gmukh@phy.iitb.ac.in*

**Abstract.** Using full potential density functional calculations within local density approximation (LDA), we report our investigation of the structural electronic properties of silicene (the graphene analogue of silicon), the strips of which has been synthesized recently on Ag(110) and Ag(100) surfaces. An assumed planar and an optimized buckled two dimensional (2D) hexagonal structures have been considered for comparisons of their electronic properties. Planar silicene shows a gapless band structure analogous to the band structure of graphene with charge carriers behaving like mass-less Dirac fermions, while the structurally optimized buckled silicene shows a small direct energy band gap of about 25 meV (at the K point of the hexagonal Brillouin zone) in its electronic structure and the charge carriers in this case behave like massive Dirac fermions. The actual band gap would be larger than this as LDA is known to underestimate the gap. The average Fermi velocity of the Dirac fermions in silicene was estimated at about half the value experimentally measured in graphene. These properties of silicene are attractive for some of the applications one envisages for graphene. Our finding of a direct band gap in silicene is something new. The results, if verified by experiments, are expected to have huge industrial impact in the silicon-based nano-electronics and nano-optics because of the possible compatibility silicene with current silicon-based micro-/nano technology.

# INTRODUCTION

Graphene, a two-dimensional crystal of C atoms covalently bonded in a two dimensional (2D) hexagonal lattice, now backed by the 2010 Nobel Prize in Physics only six years after its discovery, is currently the hottest material in the world of nano-science and nanotechnology because of its striking properties which has promising prospects for many novel applications in future nanoelectronic devices [1–6]. However, integration of graphene into the current Si-based micro-/nanotechnology and the replacement of Si electronics are tough hurdles. On the other hand, silicene (graphene-like two-dimensional (2D) hexagonal structure of Si) has attracted much attention recently both in experiments [7–9], and theory [10–14] for its expected compatibility with contemporary Si-based micro-electronics technology. The experimental realizations of the epitaxial growth of silicene strips (i.e. silicene nanoribbons) on Ag(110) and Ag(100) surfaces [7-11], have created much interest in the physics of this emerging functional material. Theoretical studies [10-15] on planar silicene (PL-Si)[10] and buckled (puckered) silicene (BL-Si) [11-13, 15] reported the graphene-like band structures of silicene.

Here, we report our comparative computational study of the electronic properties of Pl-Si and BL-Si using the density functional theory (DFT) based full potential (linearized) augmented plane wave plus local orbital (FP-(L)APW+lo) method [16] within Perdew-Zunger variant of LDA [17].

# CALCULATION METHODS

We use the elk-code [18] for our calculations. The plane wave cutoff of $|\mathbf{G+k}|_{max}$ = 9.0/$R_{mt}$ (a.u.$^{-1}$) ($R_{mt}$ is the muffin-tin radius in the unit cell) was used for the plane wave expansion of the wave function in the interstitial region. The k-point sampling [19] with a grid size of 20×20×1 was used for structural calculations and 30×30×1 for band structure and 60×60×1 for density of states (DOS) calculations. The convergence of total energy was 2.0 μeV/atom between the last two successive steps. The 2D hexagonal structures of PL-Si and BL-Si were simulated using three-dimensional hexagonal super-cells with a large value of the "$c$" ($c$ = 40 a.u.) parameter and with in-plane lattice parameter as "$a$" (= |$\mathbf{a}$| = |$\mathbf{b}$|). In BL-Si the two Si atoms are positioned at (0, 0, 0) and (2/3, 1/3, Δ/$c$) in the reduced coordinates of the 3D-super-cell; the buckling parameter Δ = 0 Å for PL-Si. The pictorial views of PL-Si and BL-Si are shown in Figure 1.

# RESULTS AND DISCUSSIONS

The ground state in-plane lattice constant of an assumed planar structure of silicene was calculated as $a_0$(PL-Si) = 3.8453 Å. For the BL-Si the optimized values of $a$ and Δ were estimated as $a_0$(BL-Si) = 3.8081 Å, Δ = 0.435 Å. These values are in agreement with reported results [10-15]. As seen in Figure 2, BL-Si is energetically more stable than PL-Si as its ground state energy $E_0$(BL-Si) is about 27 meV lower than the energy $E_0$(PL-Si). However, as seen in Figure 3 (which depicts the relative stability of silicene with respect to $\Delta$ and $a$) and Figure 4 (which depicts the variation of $\Delta$ with $a$), our assumed PL-Si structure with $a_0$(PL-Si)= 3.8453 Å should energetically tend to be buckled with a buckling parameter close to the value of 0.388 Å. To verify the validity of this approach of determining the buckling of 2D-crystals in general, we tested it in the case of graphene as depicted in Figure 5 which shows $\Delta$ = 0.00 Å for graphene (C) at our calculated ground state lattice parameter $a_0$(C) = 2.445 Å. Thus, we theoretically establish that unlike graphene, the hypothesis of graphene-like planar structure of silicene is untenable. Our calculated data on the value of $\Delta$, which minimizes the total energy for different values of $a$, best fit with a linear equation given in Figure 4. This equation may be used to estimate/predict the value of $\Delta$ corresponding to a particular $a$ value of silicene.

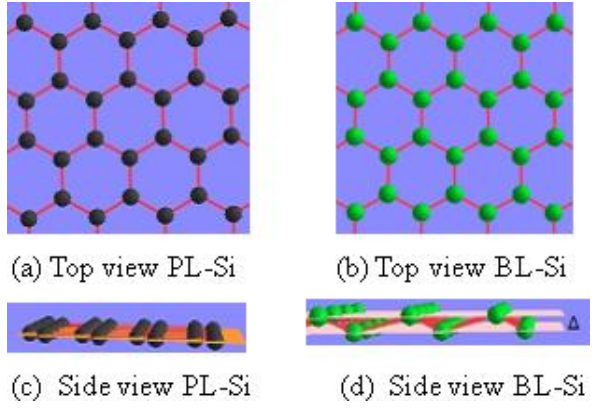

**Figure 1** Top and side views of PL-Si and BL-Si; the top-down views of PL-Si in (a) and BL-Si in (b); (c) and (d) show the side views of PL-Si and BL-Si respectively. The six Si atoms in the hexagon are in the same plane in PL-Si. But in case of BL-Si, alternate atoms are positioned in two different parallel planes; the buckling parameter $\Delta$ is the perpendicular distance between these two planes.

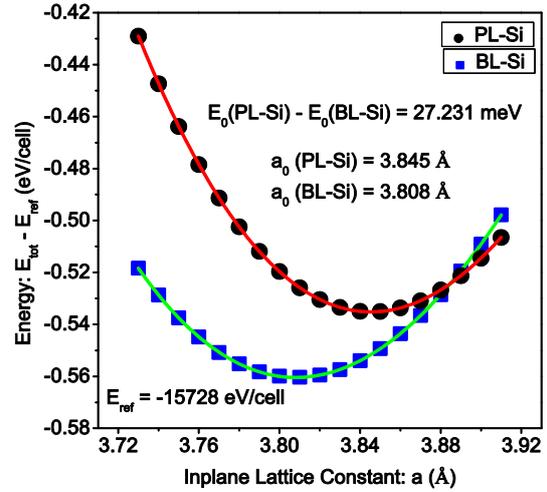

**Figure 2** Comparison of energy landscapes of PL-Si($\Delta$ = 0 Å) and BL-Si($\Delta$ = 0.435 Å).

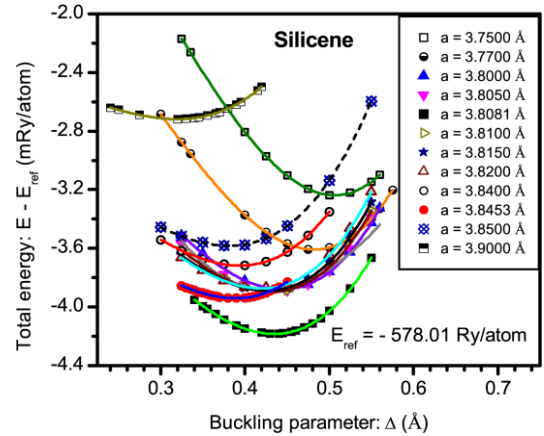

**Figure 3** Energy landscapes of silicene depicting its relative stability with respect to $\Delta$ and $a$.

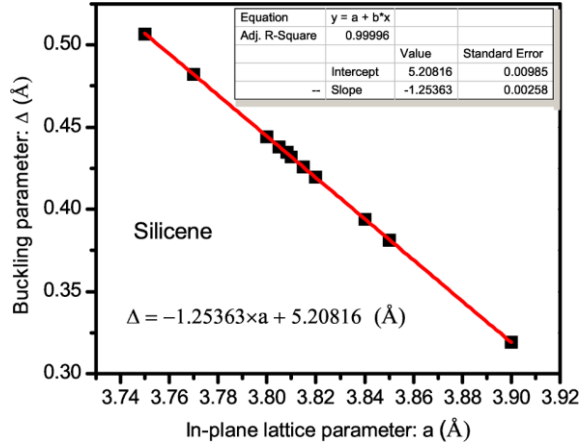

**Figure 4** Variation of buckling parameter Δ of silicene with its in-plane lattice parameter *a*.

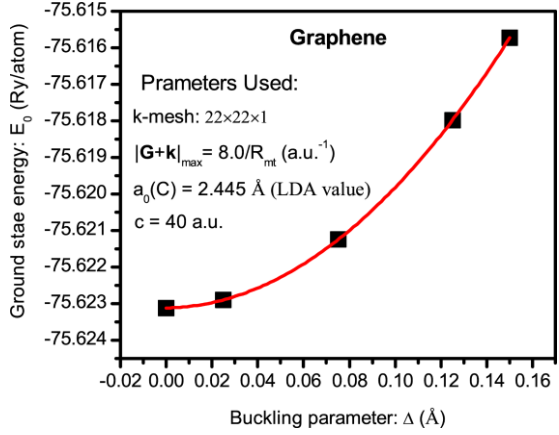

Figure 5 Probing the buckling in graphene.

The calculated band structures and total density of states (DOS) are shown in Figure 6. On the eV scale of Figure 6(a), the so called Dirac cone around the K (K′) point appears to be preserved both for PL-Si and BL-Si. However, on investigating in a finer energy scale around that point in the Brillouin Zone (BZ) (see the Figures 6(b)-(c)), we found that

(A) PL-Si has graphene-like bands with energy dispersion around the K point (also called Dirac point) being linear, i.e.,

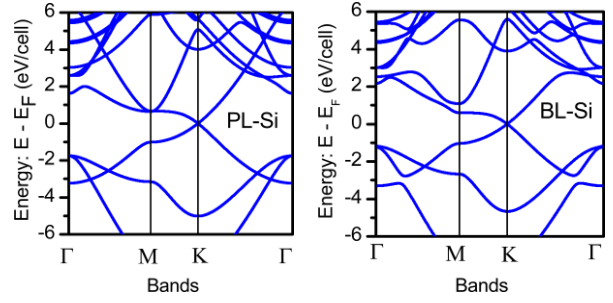

(a) Bands of PL-Si and BL-Si

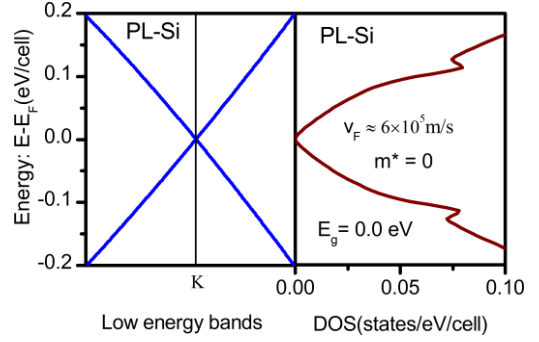

(b) Bands and DOS of PL-Si.

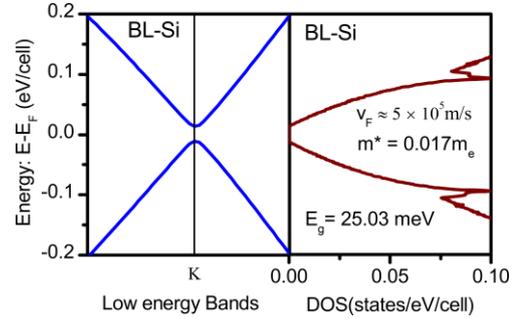

(c) Bands and DOS of BL-Si.

**Figure 6** (a) Energy bands of PL-Si and BL-Si. within LDA in the eV scale. Low energy bands and DOS of PL-Si and BL-Si are depicted in (b) and (c) respectively.

low-energy electrons and holes thus mimicking mass-less Dirac fermions' behavior [3-6],

$$E_\pm = E_F \pm \hbar k v_F \qquad (1)$$

where (ℏ$k$) is the momentum $E_F$ is the Fermi energy, $v_F$ is the Fermi velocity of the charge carriers;

(B) BL-Si has a direct band gap of 25.03 meV (actual band gap would be more than this as LDA is known to underestimate the gap) and very low energy dispersion around the Dirac points is quadratic although low energy dispersion is linear.

The quasi-particles in BL-Si mimic the massive Dirac fermions' behavior of bilayer graphene [3-6]. In fact, BL-Si which belongs to the point group rotation symmetry $C_3$, is essentially a bilayer of two triangular sub-lattices. However, near the K point of BZ, the number of band lines in BL-Si is half the number of band lines found in the case of bi-layer graphene. Further, we see many similarities in the bands of PL-Si and BL-Si except that specific degeneracies split due to lowering of point group rotation symmetry from C6 in PL geometry to $C_3$ in BL geometry. It is important to note that while searching for band gap in meV scale, choice of a finer energy scale is necessary for revealing the gap; otherwise one would erroneously get a gap-less picture as we found earlier [15] and probably obtained by others in their works [11–13] on silicene, where coarse energy scales have been used. In fact, many authors have used finer energy scales to reveal the band gap in the meV scale, in their study of other 2D materials [20-23].

The average Fermi velocity $v_F$ of quasi-particles in PL-Si and BL-Si are respectively estimated to be ≈ $0.6 \times 10^6$ m/s and ≈ $0.5 \times 10^6$ m/s, by fitting the low energy bands near the K point of the Brillouin zone to Eq. (1). The $v_F$ of silicene is about half the $v_F$ value reported in graphene [3-6], i.e., ≈ $1 \times 10^6$ m/s. The effective mass of so-called massive Dirac fermions in BL-Si was calculated as $0.017 m_e$ ($m_e$ = rest mass of electron), which is smaller than the effective mass of Dirac fermions in bilayer graphene, i.e., $0.03 m_e$ [24].

## CONCLUSIONS

We have studied and compared the electronic structures of an assumed PL-Si and relatively more stable BL-Si using the first principles full-potential DFT calculations. Both the structures are interesting in their peculiar properties. BL-Si was found energetically more stable then PL-Si. Our prediction (within LDA) of a small but finite direct band gap of BL-Si is significant. The band gap energy of BL-Si turns out to be comparable to $k_B T$ at room temperature and hence testable in experiments in future. The average Fermi velocity of the Dirac fermions in silicene was estimated at about half the value experimentally measured in graphene and the average effective mass of Dirac fermions in BL-Si is about half the corresponding value reported for bilayer graphene. These properties of silicene are attractive for some of the applications one envisages for graphene. The results, if verified by experiments, are expected to have huge industrial impact in the silicon-based nano-electronics and nano-optics because of the expected compatibility silicene with current silicon-based micro-/nano technology.

## REFERENCES

[1]. A. K. Geim, *Rev. Mod. Phys.* **83**, 851 (2011).

[2]. K. S. Novoselov, *Rev. Mod. Phys.* **83**, 837 (2011).

[3] A. K. Geim, and, K. S. Novoselov, *Nat. Mater.* **6**, 183 (2007).

[4]. A. H. Castro Neto, F. Guinea, N. M. R. Peres, K. S. Novoselov, and, A. K. Geim, *Rev. Mod. Phys.* **81**, 109 (2009).

[5]. A. K. Geim, *Science* **324**, 1530 (2009).

[6]. D.S.L. Abergel, V. Apalkov, et. al., *Adv. in Phys.* **59**, 261(2010).

[7]. A. Kara, et. al, *J. Supercon. Nov. Magn.* **22**, 259 (2009).

[8]. B. Aufray, et. al., *Appl. Phys. Lett.* 96, 183102 (2010).

[9]. D.E. Padova, C. Quaresima, et. al., *Appl. Phys. Lett.* 98, 081909 (2011).

[10]. S. Lebégue, O. Eriksson, *Phys. Rev. B* 79, 115409 (2009).

[11]. H. Şahin, S. Cahangirov, M. Topsakal, E. Bekaroglu, E. Aktürk, R.T. Senger, S. Ciraci, *Phys. Rev. B* **80**, 155453 (2009).

[12]. S. Cahangirov, M. Topsakal, E. Aktürk, H. Şahin, S. Ciraci, *Phys. Rev. Lett*. 102, 236804 (2009).

[13]. S. Wang, *J. Phys. Soc. of Jpn*. **79**, 064602 (2010).

[14]. M. Houssa, G. Pourtois, V.V. Afanas'ev, A. Stesmans, *Appl. Phys. Lett*. 97, 112106 (2010).

[15]. H. Behera, G. Mukhopadhyay, *AIP Conf. Proc*. **1313**, 152 (2010); [arXiv:1111.1282](arXiv:1111.1282).

[16]. E. Sjöstedt, L. Nordström and D. J. Singh, *Solid State Commun.* **114**, 15 (2000).

[17]. J. P. Perdew, and A. Zunger, *Phys. Rev. B,* **23**, 5048 (1981).

[18]. Elk is an open source code freely available at: http://elk.sourceforge.net/

[19] H. J. Monkhorst, and, J. D. Pack, *Phys. Rev. B* **13**, 5188 (1976).

[20]. S.-M. Choi, S.-H. Jhi, and, Y.-W. Son, *Nano Lett.* **10**, 3486 (2010).

[21]. G. Giovannetti, P. A. Khomyakov, G. Brooks, P. J. Kelly, and, J. van den Brink, *Phys. Rev. B* **76**, 073103 (2007).

[22]. J.B. Oostinga, H.B. Heersche, X. Liu, A.F. Morpurgo, L.M.K. Vandersypen, *Nature Mater.* **7**, 151 (2008).

[23]. E.V. Castro, K.S. Novoselov, et. al., *Phys. Rev. Lett.* **99**, 216802 (2007).

[24]. E. V. Castro, N. M. R. Peres, J. M.B.L. dos Santos, F. Guinea, A.H. Castro Neto, *J. Phys.: Conf. Ser*. **129**, 012002 (2008).